\documentclass[letter]{aa} % for the letters 
\usepackage{graphicx}
\usepackage{txfonts}
\usepackage{natbib}
\usepackage{amsmath}
\usepackage{color}
\usepackage{array}
\usepackage{multirow}
\usepackage{subfig}
\usepackage{xspace}
\usepackage{xcolor}
\usepackage{url}
\usepackage{rotating}
\usepackage{pdflscape}

\newcommand{\equ}[1]{Eq.~\ref{eq:#1}}

\newcommand{\fig}[1]{Fig.~\ref{fig:#1}}

\newcommand{\tab}[1]{Table~\ref{tab:#1}}

\newcommand{\lcdm}[0]{$\Lambda$CDM\xspace}
\newcommand{\re}[0]{\ensuremath{R_\mathrm{e}}\xspace}
\newcommand{\aN}[0]{\ensuremath{a_\mathrm{N}}\xspace}
\newcommand{\rbr}[0]{\ensuremath{r_\mathrm{br}}\xspace}
\newcommand{\racc}[0]{\ensuremath{r_\mathrm{acc}}\xspace}
\newcommand{\rsh}[0]{\ensuremath{r_\mathrm{sh}}\xspace}

\begin{document}

\title{Imprint of the galactic acceleration scale on globular cluster systems} % of early-type galaxies}

\author{M. B\'{i}lek\inst{1}
\and
S. Samurovi\'c\inst{2}
\and
F. Renaud\inst{3}
}
\institute{Universit\'e de Strasbourg, CNRS, Observatoire astronomique de Strasbourg (ObAS), UMR 7550, 67000 Strasbourg, France\\
              \email{bilek@astro.unistra.fr}
\and
Astronomical Observatory, Volgina 7, 11060 Belgrade, Serbia
\and
Lund Observatory, Sölvegatan 27, Box 43, SE-221 00 Lund, Sweden
}

\date{Received ...; accepted ...}

% 5 {} token are mandatory
%%%%%%%%%%%%%%%%%%%%%%%%%%%%%%%%%%%%%%%%%%%%%%%%%%%%%%%%%%%%%%%%% 
\abstract
%%%%%%%%%%%%%%%%%%%%%%%%%%%%%%%%%%%%%%%%%%%%%%%%%%%%%%%%%%%
%%%%%%%%%%%%%%%%%%%%%%%%%%%%%%%%%%%%%%%%%%%%%%%%%%%%%%%%%%%
% context heading (optional)
{We report that the density profiles of globular cluster (GC) systems in a sample of 17 early-type galaxies (ETGs) show breaks at the radii where the gravitational acceleration exerted by the stars equals the galactic acceleration scale $a_0$ known from the radial acceleration relation or{ the modified Newtonian dynamics  (MOND)}. The match with the other characteristic radii in the galaxy is not that close. We propose possible explanations {in the frameworks of the Lambda cold dark matter (\lcdm) model and MOND.} We find tentative evidence that in {the \lcdm context}, GCs reveal not only the masses of the dark halos through the richness of the GC systems but also the concentrations through the break radii of the GC systems.
}

%%%%%%%%%%%%%%%%%%%%%%%%%%%%%%%%%%%%%%%%%%%%%%%%%%%%%%%%%%%
%%%%%%%%%%%%%%%%%%%%%%%%%%%%%%%%%%%%%%%%%%%%%%%%%%%%%%%%%%%
% aims heading (mandatory)
%{}
%%%%%%%%%%%%%%%%%%%%%%%%%%%%%%%%%%%%%%%%%%%%%%%%%%%%%%%%%%%
%%%%%%%%%%%%%%%%%%%%%%%%%%%%%%%%%%%%%%%%%%%%%%%%%%%%%%%%%%%
% methods heading (mandatory)
%{}
%%%%%%%%%%%%%%%%%%%%%%%%%%%%%%%%%%%%%%%%%%%%%%%%%%%%%%%%%%%
%%%%%%%%%%%%%%%%%%%%%%%%%%%%%%%%%%%%%%%%%%%%%%%%%%%%%%%%%%%
% results heading (mandatory)
%{}
%{}


\maketitle

%%%%%%%%%%%%%%%%%%%%%%%%%%%%%%%%%%%%%%%%%%%%%%%%%%%%%%%%%%%%%%%%%
%%%%%%%%%%%%%%%%%%%%%%%%%%%%%%%%%%%%%%%%%%%%%%%%%%%%%%%%%%%%%%%%%%%%%%%%%%%%%%%%%%%%
\section{Introduction} \label{sec:intro}
Radial surface density profiles of GC systems have traditionally been described either by a power law or by a S\'ersic profile (see, e.g., the review \citealp{brodie06}). The papers investigating the kinematics of GC systems to perform Jeans analysis prefer  instead a broken power law over the S\'ersic profile for its easier computational implementation (e.g., \citealp{samur14}). In our previous work, \citet{bil19}, dealing with kinematics of GC systems of 17 early-type
galaxies (ETGs) we fitted the \textit{\emph{volume}} density profile $\rho(r)$ by a broken power law:
\begin{equation}
\begin{aligned}
\rho(r) & = \rho_0\ r^a &\quad\textrm{ for }\quad r<r_\mathrm{br},\\
\rho(r) & = \rho_0\ r_\mathrm{br}^{a-b}\ r^b &\quad\textrm{ for }\quad r\geq r_\mathrm{br}.
\end{aligned}
\label{eq:sdprof}
\end{equation} 
 Here we report that the break radii $r_\mathrm{br}$ are very close to the radii where the gravitational accelerations generated by the stars of the galaxies equal the much discussed galactic acceleration scale $a_0$ and suggest possible explanations {in the frameworks of the Lambda cold dark matter (\lcdm) model }and modified Newtonian
dynamics (MOND).

This scale, $a_0 = 1.2\times 10^{-10}$\,m\,s$^{-1}$, is revealed most clearly by rotation curves of spiral galaxies: Newtonian dynamics requires dark matter for the explanation of the rotation curves only beyond the galactocentric radii where the gravitational acceleration predicted by Newtonian dynamics, $a_\mathrm{N}$, is lower than $a_0$.  In this weak field region, the measured accelerations of stars or gas turns out to be $\sqrt{\aN a_0}$ (see, e.g., a recent study by \citealp{mcgaugh16}). This behavior was predicted  by the MOND hypothesis before detailed rotation curves were available \citep{milg83a}. According to MOND, the laws of physics need to be updated such that dynamics becomes nonlinear in the regions of space where all accelerations are below $a_0$. Assuming the standard \lcdm framework, \citet{navarro17}  proposed that the MOND-like behavior in spiral galaxies stems from the following factors: 1) the galaxies are embedded in dark halos with Navarro-Frenk-White (NFW)  profiles \citep{nfw}, 2) the mass of the halo correlates with the baryonic mass of the galaxy, 3) the baryonic mass of a disk galaxy correlates with the scale length of its exponential profile, and 4) the rotation curves can only be observed  to about five scale lengths. How these suggestions compare to observational data has not yet been assessed quantitatively, and it has not been proven, for example, that the necessary stellar-to-halo mass relation is indeed a consequence of the \lcdm theory (see \citealp{bil19} for details).  {The acceleration scale is present in the dynamics of some, if not all, ETGs} as well{ (e.g., \citealp{durazo17, durazo18})}, although verifying this is observationally difficult (see, e.g., the reviews in \citealp{milg12} and \citealp{bil19}). Even the profiles of the stellar velocity dispersion of individual globular clusters usually become flat at the radii where $\aN \approx a_0$ \citep{scarpa03, scarpa07, scarpa10, hernandez12b,hernandez13, hernandez17}. The acceleration scale $a_0$ is reflected in several other laws, such as the Faber-Jackson relation, the baryonic Tully-Fisher relation, the Fish law, and the Freeman limit, and it even coincides with the natural acceleration scales in cosmology (see the review in \citealt{famaey12} for details).

%The same scale  emerges if we combine the physical constants relevant in cosmology to obtain a quantity with the dimension of acceleration. Denoting $\Lambda$ the cosmological constant, $c$ the speed of light, $H_0$ the Hubble constant, $G$ the gravitational constant, and $M_\mathrm{U}$ and $R_\mathrm{U}$ the mass and radius of the observable Universe, it turns out that the accelerations $a_1 = \Lambda^{1/2}c^{2}$, $a_2 = H_0c$ or $a_3 = GM_\mathrm{U}R_\mathrm{U}^{-2}$ are all equal to $a_0$ within an order of magnitude \citep{milg83a, milgcjp}. 
%
%Fish law, Faber-Jackson, BTFR, 

In the present work we report that the galactic acceleration scale is imprinted even in the number density profiles of GC systems of ETGs since they show abrupt breaks near the radii where the gravitational accelerations caused by stars equal $a_0$. Other characteristic radii in the galaxies do not match the break radii that well. We discuss the possible reasons for this observation  in  \lcdm and in MOND contexts. If larger galaxy samples confirm our observation, then the breaks in the density profiles of GC systems can be used to estimate the dark halo concentration when working in the \lcdm framework. 

\begin{table*}[t!]
        \caption{Properties of the investigated galaxies and their GC systems.}
        \label{tab:tab}
        \centering
        \begin{tabular}{lccccccccccccccc}
                \hline\hline
                Name & $d$ & $\log L$ & $\log M$  & $\re$ & $n$ & $\rbr$ & $r_\mathrm{acc,N}$ & $r_\mathrm{acc,M}$ & $\re$ & $r_\mathrm{s, s}$  & $r_\mathrm{s, f}$ & $r_\mathrm{sh, s}$ & $r_\mathrm{sh, f}$ & $R_\mathrm{RB}$   \\ 
                & [Mpc] & [$L_\sun$] & [$M_\sun$] & [kpc] & & [kpc] & [$\rbr$] & [$\rbr$] & [$\rbr$] & [$\rbr$] & [$\rbr$] & [$\rbr$] & [$\rbr$] & [$\rbr$] & \\\hline    
                
               N\,821   &    24 &  10.5 &  11.2 &   4.7 &   4.7 & 16    & 0.73  & 1.1   & 0.30  & 13    & 4.0   & 0.61  & 0.85  & 6.2  \\
N\,1023  &  11.4 &  10.5 &  11.3 &   2.7 &   4.2 & 8.3   & 1.7   & 2.5   & 0.33  & 37    & 3.8   & 1.3   & 3.2   & --   \\
N\,1399  &    20 &  10.7 &  11.5 &   3.5 &   5.6 & 43    & 0.43  & 0.63  & 0.082 & 18    & 19    & 0.15  & 0.28  & --   \\
N\,1400  &    26 &  10.4 &  11.2 &   3.4 &   4.0 & 23    & 0.49  & 0.73  & 0.15  & 6.8   & 2.7   & 0.46  & 0.60  & 0.46 \\
N\,1407  &    29 &    11 &  11.9 &   9.4 &   8.3 & 46    & 0.52  & 0.77  & 0.21  & 3.6   & 8.7   & --    & 0.34  & 0.58 \\
N\,2768  &    22 &  10.7 &  11.5 &   8.9 &   3.3 & 21    & 0.70  & 1.1   & 0.42  & 34    & 3.8   & 0.22  & 0.72  & 7.6  \\
N\,3115  &   9.7 &  10.2 &  11.0 &   4.8 &   4.4 & 9.5   & 0.85  & 1.3   & 0.50  & 8.0   & 8.3   & 0.97  & 0.95  & --   \\
N\,3377  &  11.2 &   9.9 &  10.5 &   2.9 &   5.0 & 5.6   & 0.85  & 1.3   & 0.52  & 5.1   & 5.6   & 1.4   & 1.2   & 1.9  \\
N\,4278  &    16 &  10.2 &  11.0 &   2.5 &   4.8 & 15    & 0.61  & 0.90  & 0.16  & 5.0   & 6.6   & 0.70  & 0.56  & 0.48 \\
N\,4365  &    20 &  10.7 &  11.5 &   8.5 &   5.2 & 29    & 0.54  & 0.82  & 0.29  & 27    & 17    & 0.15  & 0.42  & 1.1  \\
N\,4472  &  16.3 &  10.9 &  11.7 &   3.9 &   3.0 & 24    & 1.0   & 1.5   & 0.16  & 21    & 33    & --    & 0.44  & --   \\
N\,4486  &    16 &  10.8 &  11.6 &   5.8 &   2.9 & 15    & 1.4   & 2.0   & 0.40  & 54    & 4.3   & 0.16  & 0.42  & --   \\
N\,4494  &  17.1 &  10.4 &  11.1 &   3.7 &   3.4 & 10    & 1.0   & 1.5   & 0.36  & 11    & 3.9   & 1.0   & 1.5   & --   \\
N\,4526  &    17 &  10.4 &  11.2 &   2.7 &   3.6 & 12    & 1.0   & 1.5   & 0.22  & 13    & 6.5   & 0.92  & 0.99  & --   \\
N\,4649  &    17 &  10.8 &  11.6 &   5.1 &   3.6 & 30    & 0.69  & 1.0   & 0.17  & 25    & 11    & 0.074 & 0.53  & --   \\
N\,5128  &   4.2 &  10.5 &  11.2 &   6.2 &   4.0 & 12    & 0.93  & 1.4   & 0.54  & 18    & 22    & 0.77  & 0.65  & --   \\
N\,5846  &    25 &  10.7 &  11.6 &   8.1 &   3.9 & 37    & 0.44  & 0.68  & 0.22  & 21    & 8.5   & 0.099 & 0.30  & 0.77 \\
\hline
R17x     &    19 &  10.2 &  11.0 &   1.5 &   2.6 & 2.0   & 5.2   & 7.4   & 0.75  & 38    & 16    & 6.0   & 10    & --   \\
R17y     &    19 &  10.2 &  11.0 &   1.7 &   2.0 & 3.0   & 3.6   & 5.1   & 0.57  & 26    & 13    & 4.1   & 5.7   & --   \\
R17z     &    19 &  10.2 &  11.0 &   2.0 &   2.0 & 3.4   & 3.1   & 4.4   & 0.58  & 22    & 9.2   & 3.5   & 4.3   & --   \\

                \hline
        \end{tabular}
        \tablefoot{
                $\boldsymbol{d}$ -- Galaxy distance. 
                $\boldsymbol{L}$ -- Galaxy $B$-band luminosity.
                                                                $\boldsymbol{M}$ -- Galaxy stellar mass.
                $\boldsymbol{R_\mathrm{e}}$ -- S\'ersic effective radius.  
                $\boldsymbol{n}$ -- S\'ersic index. 
                $\boldsymbol{\rbr}$ -- Break radius of the number density profile of the GC system.
                                                                $\boldsymbol{r_\mathrm{acc,N}}$ -- Radius where the Newtonian acceleration equals $a_0$.
                                                                $\boldsymbol{r_\mathrm{acc,M}}$ -- Radius where the MOND acceleration equals $a_0$.
                                                                $\boldsymbol{r_\mathrm{s, s}}$ -- Scale radius of a NFW halo following the scaling relations calculated from the stellar mass. 
                                                                $\boldsymbol{r_\mathrm{s, f}}$ -- Scale radius of a NFW halo obtained by \citet{bil19} from fitting the GC kinematics.
                                                                $\boldsymbol{r_\mathrm{sh, s}}$ -- Radius at which the acceleration originating from the stars and the dark halo are equal. The halo parameters were estimated from the scaling relations. A dash means that the acceleration exerted by dark matter is stronger than the acceleration exerted by the stars at all radii. 
                                                                $\boldsymbol{r_\mathrm{sh, f}}$ -- The same, but the halo parameters were estimated by fitting the GC kinematics.
                                                                $\boldsymbol{R_\mathrm{RB}}$ -- Radius where surface densities of the red and blue GCs are equal. A dash means  missing data.
                  }
\end{table*}

\begin{figure}
        \resizebox{\hsize}{!}{\includegraphics{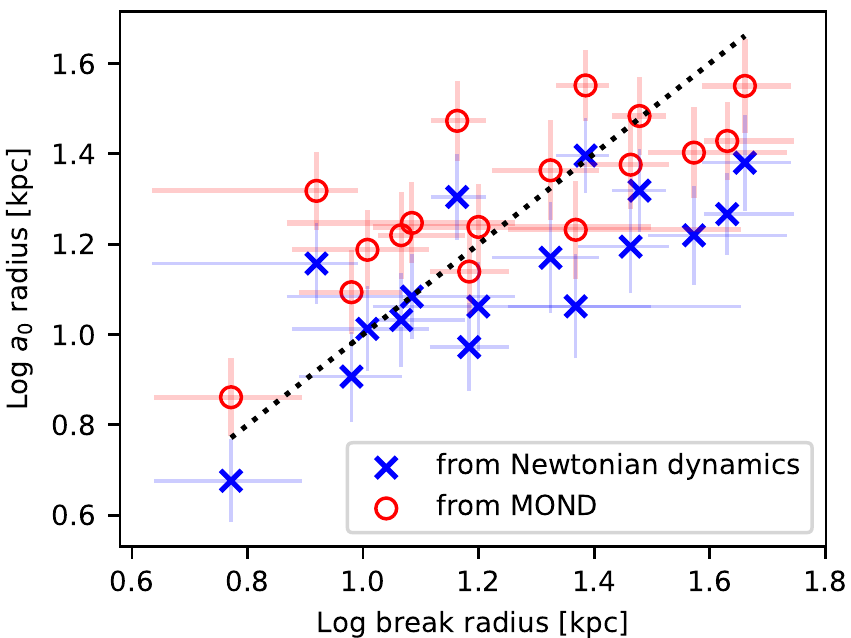}}
        \caption{ Demonstration that the break radii of GC systems are nearly equal to the radii where the gravitational accelerations equal the galactic acceleration scale $a_0$. The gravitational acceleration was calculated from the distribution of stars and  either the Newtonian gravity (crosses) or MOND (circles). The dotted line marks the one-to-one relation.} 
        \label{fig:plot}
\end{figure}

\section{Break radii of GC systems}
%table name, d, L, M, Re, n, xbr, r_aN/xbr, r_aM/xbr, r_s/xbr, Re/xbr, rRB?
We use the data set presented in \citet{bil19}. Briefly, in that work we collected archival data on GC systems of 17 nearby  ETGs, most of which  have over 100 archival GC radial velocity measurements available. We included in our sample the GCs of a galaxy formed in the cosmological \lcdm hydrodynamical simulation reported in \citet{renaud17}. We investigated the gravitational fields of these objects through the Jeans equation. This requires obtaining the number density profiles of the GC systems. {We found that these profiles can be nicely described by a broken power law, and we provided fits to the profiles} (see Sect.~3 of  \citealp{bil19}).  {Additional details on the fitting and the error bars are provided in the   Appendix~\ref{sec:fits}.} The density profiles become steeper beyond the breaks.  The calculated break radii \rbr are reproduced in \tab{tab} along with other basic characteristics of the galaxies, which we adopt hereafter (the objects R17x, y, and z are three perpendicular projections of the simulated galaxy). We checked in \citet{bil19} that the breaks are not due to the incompleteness of the survey because they correlate with the mass of the galaxy. Similar behavior was noted for the effective radii of GC systems \citep{forbes17}.

In \tab{tab}, we compare the break radii \rbr to other characteristic lengths in the galaxy. It is striking that the break radii are always very close to the radius $r_\mathrm{acc,N}$ where the gravitational acceleration generated by the stars according to Newtonian gravity equals $a_0$ (their average ratio is $0.8\pm0.3$, not counting the artificial galaxies). The same is true for the radius $r_\mathrm{acc,M}$ defined analogously for the MOND gravity (average ratio  $1.2\pm0.5$). The MOND gravitational field was calculated using Eq.~(4) in \citet{mcgaugh16}. In contrast, the radii $r_\mathrm{acc,N}$ and $r_\mathrm{acc,M}$ for the simulated galaxy differ from \rbr by a factor of a few. Since $r_\mathrm{acc,N}$ and $r_\mathrm{acc,M}$ are similar, we often refer to them in the following just as  \racc. The comparison of the \racc and \rbr radii is presented graphically in \fig{plot}\footnote{{The error bars were calculated using the standard formula $\left[\Delta f(x_1, x_2, \ldots)\right]^2 = \sum (\partial f/\partial x_i \Delta x_i)^2$, where the {uncertainties on the mass-to-light ratio and galaxy distance were taken from \citet{bil19} and those on} \rbr from \tab{fits}.} }. The break radii  \rbr are apparently different from the projected half-light radii of the galaxies \re (average ratio $0.3\pm0.1$).  We considered two ways of estimating the scale radii of the dark NFW halos. The halo scale radius  $r_\mathrm{s, s}$ was obtained by combining the stellar-to-halo mass relation \citep{behroozi13} and the halo mass-concentration relation \citep{diemer15}. The halo scale radius  $r_\mathrm{s, f}$ was fitted in \citet{bil19} directly on the  kinematics of the GCs studied here. \tab{tab} shows that the both types of   scale radii are many times greater than \rbr. In spiral galaxies, where gravitational fields can be investigated precisely by rotation curves, the \racc radius proved to bound the region beyond which Newtonian dynamics requires dark matter. This led us to hypothesize that, in the \lcdm context, \racc is the radius where the acceleration caused by stars equals the acceleration caused by the dark halo, \rsh. Here we again consider $r_\mathrm{sh, s}$ for the halos estimated from the scaling relations and $r_\mathrm{sh, f}$ for the fitted halos. Particularly $r_\mathrm{sh, f}$ are not that far from \rbr given that{ the uncertainty on $r_\mathrm{sh, f}$} is around a factor of two (Table~5 of \citealp{bil19}). {The average ratio of  $r_\mathrm{sh, f}$ to \rbr is} $0.8\pm0.7$. Again, the simulated galaxies differ substantially. One might further suspect that the breaks in the number density profiles are caused by different density profiles of the blue and red GCs because the blue GCs are known to be generally more extended than the red GCs (e.g., \citealp{brodie06}). Then a break  in the surface density would occur at the radius $R_\mathrm{RB}$  where the surface densities of the two types of GCs are equal. We found these radii  for several galaxies in the surface density plots by \citet{pota13}. \tab{tab} shows that $R_\mathrm{RB}$ and $\rbr$ differ substantially in some cases. We have checked that the acceleration profiles of the galaxies do not have local maxima near the break radii, regardless of whether with Newtonian gravity and the two considered types of dark halos or with MOND. These profiles actually have maxima only in the galaxy centers.

\begin{figure}
        \resizebox{\hsize}{!}{\includegraphics{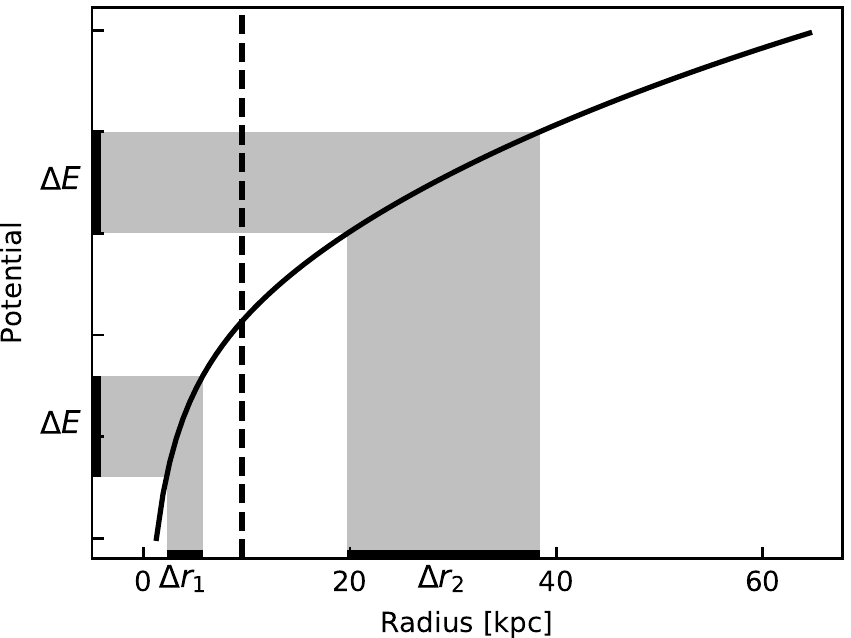}}
        \caption{Illustration of a possible explanation of the breaks of the density profiles of GC systems in the \lcdm framework proposed in the text. The solid line shows the gravitational potential of NGC\,3115 obtained by fitting the GC kinematics \citep{bil19}. The vertical dashed line marks the observed break radius of the GC system that virtually matches the \rsh radius.  A similar explanation works for MOND as well, but the break occurs at the radius \racc. The potential is given in arbitrary units.}% that matches the break radii better for our galaxies with the current data (\tab{tab}). } 
        \label{fig:illust}
\end{figure}

\section{Discussion and conclusions}
We found that the breaks in the profiles of GC number density are very close to the radius \racc where the gravitational acceleration caused by stars equals the galactic acceleration scale $a_0$. Our sample probed this for galaxy masses ranging over one order of magnitude. But what could be the reason for this? {The aim of our paper is primarily to report the observational finding, but here we present our ideas about the theoretical explanation.}

Let us recall that in the standard scenario of formation of GC systems in {the \lcdm context}, most red GCs are formed in a galaxy together with its stars, while most blue GCs are brought by the accreted satellites \citep{cote98,tonini13,renaud17}. This hierarchical buildup of galaxies was proposed to explain the well-known correlation between the number of GCs and the mass of the halo \citep{harris15,burket19}\footnote{However, as  discussed in \citet{burket19} the available detailed models of this process do not match all the observational and theoretical constraints.}. Our \lcdm explanation of the break radii is based on the hierarchical buildup and the assumption that the gravitational potential has different behavior under and beyond \rsh. Inside this radius the gravitational field is mostly determined by the concentrated baryonic component, while beyond \rsh the potential is mostly given by the extended dark halo. As an illustration, we plot in \fig{illust} the gravitational potential of NGC\,3115, which exhibits a good match between \rsh and \rbr. {Let us consider two satellites} S1 and S2 that are approaching the center of the galaxy; they  have all  the same properties, but the radial velocity of S2 is greater.  Both satellites are disrupted by tidal forces at the same galactocentric radius, but the average energies of the released GCs are higher for   satellite S2. At the same time, we expect that the energy span $\Delta E$ of the GC systems of the two satellites originating in the internal velocity dispersions of the GC systems will be nearly the same. Figure~\ref{fig:illust} illustrates that the span of radii occupied by the GCs from the slower satellite S1, $\Delta r_1$, is smaller for a suitable choice of  impact velocities than that of the GCs from the faster satellite S2, $\Delta r_2$ (for simplicity we assumed that the satellites approach the main galaxy radially) This is so because the slope of the gravitational potential becomes shallower beyond  \rsh.  With this simplified treatment we then obtain a break in the density profiles of the GC system. Another motivation for the breaks is provided by the Jeans equation. It connects the gravitational field with the density profile of the tracers, the velocity dispersion, and the anisotropy parameter. A break in the profile of the gravitational acceleration around \rsh then has to have a counterpart among the profiles of the other quantities appearing in the Jeans equation. 

We should keep in mind that this explanation leaves some questions open. First, the simulated galaxies have their \rbr radii different several times from the \rsh radii and even from the \racc radii. It is not clear at the moment whether this occurs because of some inadequacy of the particular simulation (e.g., because it aimed to represent a spiral galaxy at a redshift of 0.5) or if it points to a more general problem of the \lcdm galaxy formation theory. Next, the two types of \rsh listed in \tab{tab} deviate   from the \rbr radii by a factor of a few and we actually chose the galaxy NGC\,3115 for our illustration because the proposed explanation works well for it. These concerns might simply be explained by the uncertainties of the fitted dark halo parameter; for the scale radii, they are around a factor of two \citep{bil19}. Finally, as the galaxy accretes the GCs, its mass increases as well, which changes its gravitational potential and the \rsh radius. This could lead to a profile of the density of GCs that does not exhibit the observed sharp break. It is also not clear whether the match $\rbr\approx 3\re$ has any deeper meaning. Investigating these issues is beyond the scope of this paper.  

Nevertheless, if it turns out that \rbr and \rsh are indeed equal, the consequence is intriguing: bare imaging of a GC system of a galaxy enables us to  estimate both mass and concentration of its NFW halo. In particular, once we estimate the mass of the halo from the number of GCs, the halo scale radius can be calculated from the condition that the gravitational acceleration caused by the stars equals the acceleration caused by the halo at \rbr. This would allow us to investigate the gravitational fields of objects where it is currently difficult to do so spectroscopically, such as ETGs or low surface brightness objects similar to the GC-rich ultra diffuse galaxy Dragonfly\,44 \citep{vandokkum16}.

In MOND, the formation of GC systems is little explored. The above explanation of the breaks in GC density profiles based on a break in gravitational potential also applies, with the difference that the breaks in the gravitational potentials occur at \racc simply because of the change between the two MOND regimes, {as already pointed out by \citet{hernandez13b}}.  In MOND, galaxy mergers are not expected to be as frequent as in {the \lcdm context} \citep{nipoti07,tiret08,combtir10,kroupacjp}, but GCs can be transferred between galaxies even during nonmerging galaxy encounters \citep{bekki03,bil18}.  {Nevertheless,} a diversity of events in the life of a galaxy are expected  to strengthen the break in  the GC system in the MOND framework because the radius \racc separates the inner region governed by the linear Newtonian dynamics from the outer region following the nonlinear deep-MOND dynamics. For example, the value of the gravitational constant in the Chandrasekhar formula for dynamical friction is effectively increased in the region beyond \racc  for the case of GCs (\citealp{nipoti08}; this is not necessarily true for interacting galaxies). We would then expect the outer GCs to spiral into the Newtonian region making   the profile steeper beyond  \racc and shallower below, in agreement with observations\footnote{This consideration actually led MB to compare the break radii with \racc, and after that with all the other characteristic radii. The breaks in the GC system profiles at \racc can thus be taken as a MOND prediction.}. The MOND nonlinearity causes the external field effect \citep{milg83a, BM84, milgmondlaws}. When a galaxy that was originally isolated enters a galaxy cluster where the strength of the gravitational field is greater than the internal strength of the gravitational field of the galaxy, the external field effect causes the internal gravity of the galaxy to reduce beyond \racc (as if dark matter were removed from the galaxy in the Newtonian view).  The GCs outside of \racc would then expand lowering the density of the GC system and causing a break. Similarly, when two galaxies interact, their gravitational fields do not add linearly in the low-acceleration regions of the galaxies, but they do in the high-acceleration regions. The tidal forces then affect the GCs differently in the inner and in the outer part. Even just a change in the stellar mass $M$ of a galaxy has a different response in the two regions since the attractive force is proportional to $M$ in the inner Newtonian region and to $\sqrt{M}$ in the outer deep-MOND region. {Interestingly, \citet{milg84} calculated that the density of a self-gravitating isothermal sphere is proportional to $r^{-\alpha_\infty}$ at large radii, where $3.5 \lesssim \alpha_\infty \lesssim 4.5$, which agrees with most of our fits of the outer slope $b$ (see \tab{fits}).}

These theoretical considerations obviously need confirmation by detailed calculations. It is also desirable to verify observationally for a wider variety of galaxy masses our result that GC systems have breaks in their density profiles at the \racc radius. The reported finding provides an interesting constraint on galaxy formation and a new element to the missing mass discussion.

\begin{acknowledgements}
We acknowledge Duncan Forbes for the suggestion that the breaks in the profiles of GC systems reflect the blue and red GC subpopulations. 

We thank Xavier Hernandez for an insightful referee report.

SS acknowledges the support from the Ministry of Education, Science and Technological Development of the Republic of Serbia through project no.~176021 ``Visible and Invisible Matter in Nearby Galaxies: Theory and Observations.''
        
        FR acknowledges support from the Knut and Alice Wallenberg Foundation.

\end{acknowledgements}

\bibliographystyle{aa}
\bibliography{citace}

\begin{appendix}
\section{Fits of radial number density profiles the GC systems}
\label{sec:fits}
\begin{table*}[b!]
        \caption{Properties of the investigated galaxies and their GC systems.}
        \label{tab:fits}
        \centering
        \begin{tabular}{lccccccccccccccc}
                \hline\hline
                Name &  $\rho_0$  & \rbr  & $a$ & $b$   \\ 
                & [arcmin$^{-3}$] & [arcmin] & & \\\hline    
                
               N\,821     & 1.2$^{+0.3}_{-0.4}$            & 2.3$^{+2}_{-0.8}$              & -1.6$^{+1}_{-0.6}$             & -3.4$^{+0.6}_{-20}$            \\
N\,1023    & 2.0$^{+0.4}_{-0.4}$            & 2.5$^{+0.4}_{-1}$              & -1.9$^{+0.5}_{-0.4}$           & -3.8$^{+0.5}_{-0.4}$           \\
N\,1399    & 1.3$^{+2}_{-0.5}$              & 7.3$^{+2}_{-0.4}$              & -1.1$^{+0.3}_{-0.6}$           & -4.7$^{+0.2}_{-1}$             \\
N\,1400    & 1.5$^{+0.3}_{-0.3}$            & 3.1$^{+3}_{-0.6}$              & -2.4$^{+0.5}_{-0.4}$           & -4.5$^{+0.6}_{-0.8}$           \\
N\,1407    & 1.6$^{+0.5}_{-0.4}$            & 5.4$^{+0.8}_{-0.6}$            & -1.4$^{+0.3}_{-0.2}$           & -4.0$^{+0.3}_{-0.4}$           \\
N\,2768    & 2.1$^{+0.3}_{-0.3}$            & 3.3$^{+0.6}_{-0.6}$            & -2.2$^{+0.3}_{-0.2}$           & -5.0$^{+0.7}_{-0.9}$           \\
N\,3115    & 1.5$^{+0.6}_{-0.6}$            & 3.4$^{+0.7}_{-0.6}$            & -1.5$^{+0.6}_{-0.4}$           & -4.3$^{+0.5}_{-0.9}$           \\
N\,3377    & 2.1$^{+0.5}_{-0.5}$            & 1.8$^{+0.6}_{-0.5}$            & -1.5$^{+1}_{-0.6}$             & -3.5$^{+0.2}_{-0.3}$           \\
N\,4278    & 2.9$^{+0.5}_{-0.5}$            & 3.3$^{+0.4}_{-0.4}$            & -1.6$^{+0.3}_{-0.2}$           & -4.1$^{+0.3}_{-0.4}$           \\
N\,4365    & 2.7$^{+0.4}_{-0.4}$            & 5.0$^{+0.7}_{-0.9}$            & -2.0$^{+0.2}_{-0.2}$           & -4.7$^{+0.6}_{-0.6}$           \\
N\,4472    & 0.5$^{+0.3}_{-0.2}$            & 5.1$^{+0.4}_{-0.5}$            & -0.5$^{+0.6}_{-0.4}$           & -5.4$^{+0.7}_{-0.9}$           \\
N\,4486    & 0.9$^{+0.6}_{-0.4}$            & 3.1$^{+0.3}_{-0.2}$            & -0.1$^{+0.7}_{-0.6}$           & -3.37$^{+0.07}_{-0.08}$        \\
N\,4494    & 1.9$^{+0.4}_{-0.4}$            & 2.0$^{+0.6}_{-0.5}$            & -1.4$^{+1}_{-0.5}$             & -4.1$^{+0.4}_{-0.4}$           \\
N\,4526    & 1.8$^{+0.4}_{-0.5}$            & 2$^{+1}_{-1}$                  & -2.0$^{+0.7}_{-0.5}$           & -3.4$^{+0.4}_{-0.6}$           \\
N\,4649    & 3.2$^{+0.4}_{-0.4}$            & 6.1$^{+0.5}_{-0.5}$            & -1.8$^{+0.1}_{-0.1}$           & -5.6$^{+0.3}_{-0.4}$           \\
N\,5128    & 0.11$^{+0.3}_{-0.06}$          & 9.5$^{+3}_{-0.6}$              & -0.4$^{+0.4}_{-0.7}$           & -4.4$^{+0.1}_{-0.1}$           \\
N\,5846    & 1.1$^{+0.4}_{-0.3}$            & 5.1$^{+2}_{-0.7}$              & -1.4$^{+0.3}_{-0.3}$           & -4.2$^{+0.6}_{-10}$            \\
R17x       & 120$^{+200}_{-70}$             & 0.36$^{+0.03}_{-0.03}$         & -0.5$^{+0.1}_{-0.1}$           & -3.8$^{+0.1}_{-0.2}$           \\
R17y       & 20$^{+10}_{-6}$                & 0.54$^{+0.05}_{-0.05}$         & -1.61$^{+0.1}_{-0.09}$         & -3.8$^{+0.2}_{-0.2}$           \\
R17z       & 17$^{+7}_{-4}$                 & 0.62$^{+0.05}_{-0.06}$         & -1.6$^{+0.1}_{-0.1}$           & -3.9$^{+0.2}_{-0.2}$           \\

                \hline
        \end{tabular}
        \tablefoot{Best-fit parameters of the radial volume density profiles of the GC systems in \equ{sdprof}.
                $\boldsymbol{\rho_0}$ -- Density of GCs at the galactocentric distance of 1\arcmin.
                $\boldsymbol{\rbr}$ -- Break radius.
                                                                $\boldsymbol{a}$ -- Inner slope of the radial  density profile.
                $\boldsymbol{b}$ -- Outer slope of the radial  density profile. 
                  }
\end{table*}

We list in \tab{fits} the best-fit parameters in \equ{sdprof} to the GC systems we investigated. Compared to \citet{bil19}, we additionally state the $1\,\sigma$ uncertainty limits. The procedure of fitting is described in the original paper. Briefly, the GCs are divided into several annuli centered on the host galaxy such that every annulus contains the same number of GCs, $N$. The surface density is estimated in the middle radius of every annulus by dividing the number of GCs by the surface of the annulus, $S$. The error is estimated as $\sqrt{N}/S$ assuming a Poisson statistics. Then the resulting data points are fitted in the usual least-squares way. In \citet{bil19}, we had to convert the analytic volume density to surface density numerically during the fitting, which is computationally demanding, but we   found an analytic expression. It is rather complex, but easy to determine once we realize that the integral appearing in the Abel transform of \equ{sdprof}

\begin{equation}
\int \frac{r^\alpha}{\sqrt{r^2-R^2}} \mathrm{d}r = R^\alpha u\, _2F_1\!\left(\frac{1}{2}, -\frac{\alpha-1}{2}; \frac{3}{2}; -u^2\right) + const.,
\label{eq:a1}
\end{equation}
where $r$ means the real radius, $R$ the projected radius, $u = \sqrt{\left(r/R\right)^2-1}$, and $_2F_1$ is the Gaussian hypergeometric function. We note that   via this approach we obtain the parameters of the volume density profile directly by fitting the projected density profile. The errors of the fitted parameters were estimated in the way described in Sect.~5 of \citet{bil19}: denoting the vector of free parameters as $p = (p_1,p_2,\ldots)$, we estimated the uncertainty limits of the parameter $p_i$ by minimizing or maximizing $p_i$ over the region of the parameter space where the logarithm of the likelihood of the corresponding model does not differ from the logarithm of the likelihood of the best-fit model by more than 0.5. The likelihood  reads
\begin{equation}
\ln \mathcal{L}(p) = \sum_i -0.5\ln\left(2\pi\right)-\ln\left(\Delta\Sigma_i\right)-\frac{\left[\Sigma_\mathrm{m}(r_i, p)-\Sigma_i\right]^2}{2\Delta\Sigma_i^2},
\end{equation}
where $r_i = (r_{i,\mathrm{max}}+r_{i,\mathrm{min}})/2$ denotes the middle radius of an annulus bounded by the radii $r_{i,\mathrm{max}}$ and $r_{i,\mathrm{min}}$, $\Sigma_i$ and $\Delta\Sigma_i$ are respectively the observed surface density at $r_i$ and its uncertainty, and $\Sigma_\mathrm{m}(r_i,p)$ stands for the modeled surface density at the radius $r_i$ corresponding to the parameter vector $p$.

\end{appendix}

\end{document}